# Electronic Origins of Large Thermoelectric Power Factor of LaOBiS$_{2-x}$Se$_x$

Atsuhiro Nishida[1], Hirotaka Nishiate[2], Chul-Ho Lee[2], Osuke Miura[1], Yoshikazu Mizuguchi[1]*

[1]*Department of Electrical and Electronic Engineering, Tokyo Metropolitan University, 1-1, Minami-osawa, Hachioji 192-0397, Japan*
[2] *National Institute of Advanced Industrial Science and Technology (AIST), 1-1-1 Umezono, Tsukuba, Ibaraki 305-8568 Japan.*

We examined the electrical transport properties of densified LaOBiS$_{2-x}$Se$_x$, which constitutes a new family of thermoelectric materials. The power factor increased with increasing concentration of Se, i.e., Se substitution led to an enhanced electrical conductivity, without suppression of the Seebeck coefficient. Hall measurements indicated that the low electrical resistivity resulted from increases in the carrier mobility, and the decrease in carrier concentration led to large absolute values of the Seebeck coefficient of the system.

LaOBiCh$_2$ (Ch: S, Se) and related Bi-Ch layered compounds have attracted significant attention as new thermoelectric materials [1–4]. The crystal structure of LaOBiCh$_2$ consists of alternating stacks of an electrically conducting BiCh$_2$ bilayer and an insulating La$_2$O$_2$ layer. We recently obtained a high dimensionless figure-of-merit of $ZT$ = 0.36 (at ~650 K) for LaOBiSSe [5]. In contrast to the typically observed trend, the electrical resistivity of LaOBiS$_{2-x}$Se$_x$ was suppressed and the absolute value of the Seebeck coefficient was increased by Se substitution [2]. We therefore sought to clarify the electronic origins of the Se-substitution-induced enhancement of the power factor ($PF = S^2/\rho$) of LaOBiS$_{2-x}$Se$_x$; this was done by determining the electrical resistivity ($\rho$), Seebeck coefficient ($S$), carrier concentration ($n$), and mobility ($\mu$) of the system.

Polycrystalline samples of LaOBiS$_{2-x}$Se$_x$ ($x$ = 0, 0.4, 0.6, 0.8, and 1) were prepared using the solid-state-reaction method described in Ref. 2. The samples were densified by hot-pressing (HP) under a uniaxial pressure of ~50 MPa. The resulting LaOBiS$_{2-x}$Se$_x$ powders were sealed in a 15-mm-diameter graphite capsule during HP annealing. The powders were annealed for 1 h at 700 ºC and 800 ºC for $x$ = 0.4–1 and $x$ = 0, respectively. This annealing yielded densified samples with relative densities higher than 97%. The corresponding $\rho$ and $S$ were measured from room temperature to ~650 K, by using the four-terminal method. In addition, Hall measurements were performed in order to estimate the $n$ and $\mu$ at room temperature, by using the van-der-Pauw method at applied fields of ~0.5, 1, 1.5, and 2 T. These measurements were typically conducted on 10×10 mm-square, 1-mm-thick samples.

Figures 1(a) shows that the $\rho$ of the LaOBiS$_{2-x}$Se$_x$ samples increases with increasing temperature and, indicative of increased electrical conductivity with doping, decreases with increasing Se concentration ($x$). However, at elevated temperatures, the data for $x = 0$ deviate from this tendency. We assume that the low $\rho$ for $x = 0$ at elevated temperatures, results from the sample being annealed at a different temperature (800 ºC) from that of the other samples. A plot of the temperature-dependence of $S$ (Figure 1(b)) reveals that the $S$ of all the LaOBiS$_{2-x}$Se$_x$ samples decreases (absolute value of $S$ increases) with increasing temperature. Negative $S$ values indicate that the contributing carriers are electrons. At room temperature, the absolute value increases with increasing $x$ and is minimum at $x = 0$. The temperature-dependence of the power factor of the system is shown in Figure 1(c). As the figure shows, this factor increases with increasing $x$, and is highest for $x = 1$ over the range of temperatures examined. Power factors higher than 5.5 µW/cmK$^2$ are realized at temperatures above 450 K. Therefore, Se substitution leads to reductions in $\rho$ and increases in absolute value of $S$, which are both essential for the enhancement of $PF$ in this system.

We performed Hall measurements and estimated $n$ and $\mu$ in order to determine from an electronic viewpoint, the basis for the enhancement of $PF$. Figure 2 shows the dependence of $n$ and $\mu$ on $x$; $n$ decreases with increasing values of $x$. Increases in the absolute value of $S$ with Se substitutions may be attributed to this decrease, and $S$ is, in general, proportional to $n^{-1}$. The decrease in $n$ with increasing concentration of Se may result from changes in the electronic structure. Recently, synchrotron X-ray diffraction experiments revealed that a LaOBiS$_2$ single crystal has a monoclinic, rather than tetragonal, crystal structure [6]. In fact, we observed (via neutron diffraction) the short-range distortions of the Bi-S plane of the LaOBiS$_2$ polycrystalline sample [7]. Therefore, we assume that Se substitutions affect both the local crystal structure and electronic structure of LaOBiS$_{2-x}$Se$_x$.

In contrast to $n$, $\mu$ increases (Figure 2(b)) with increasing $x$; $\mu$ obtained at $x = 1$ is 10 times larger than that obtained at $x = 0$. The increase in $\mu$ is attributed to the chemical pressure, i.e., in the LaOBiCh$_2$-type structure, Se substitution leads to increased overlap between the Bi-6p and Ch-p orbitals, via the in-plane chemical pressure effect [8]. These overlaps may, in turn, lead to increases in the bandwidth and consequently, the enhancement of $\mu$. In general, $\rho$ can be estimated from: $\rho = 1/en\mu$, where $e$ is the elementary charge. In the present samples, the enhancement of $\mu$ is greater than the expected decrease in $n$. The increase in the electrical conductivity of LaOBiS$_{2-x}$Se$_x$ is therefore attributed to the enhancement of $\mu$.

In conclusion, we examined the dependence of *PF* ($\rho$ and *S*), *n*, and $\mu$ on the concentration of Se. The *PF* in Se-substituted LaOBiS$_{2-x}$Se$_x$ is enhanced via reductions in $\rho$ and increases in the absolute value of *S*; these are attained through a significant enhancement of $\mu$ and a decrease in *n*, respectively. The results indicate that the carrier concentration and the thermoelectric properties of the BiCh$_2$-based compounds can be controlled by Se-substitution at the conduction layers.

**Acknowledgment**

The authors thank K. Kuroki of Osaka University for fruitful discussion. This work was partly supported by Grant-in-Aid for Young Researcher (A) (25707031), Grant-in-Aid for Scientific Research on Innovative Areas (15H05886), and the TEET research fund.

*E-mail: mizugu@tmu.ac.jp

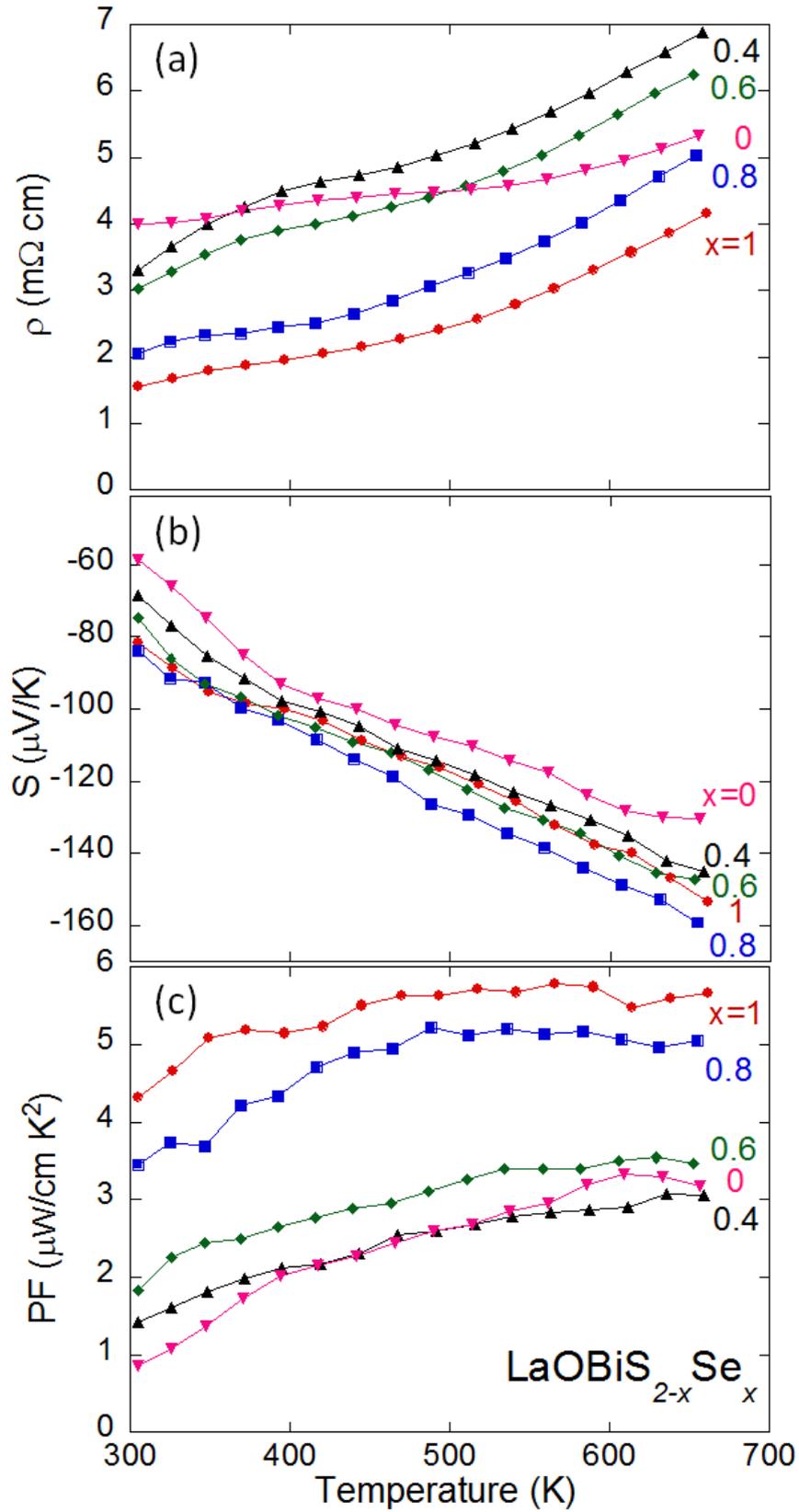

Fig. 1. Temperature-dependence of the (a) electrical resistivity ($\rho$), (b) Seebeck coefficient ($S$), and (c) power factor ($PF$) of densified LaOBiS$_{2-x}$Se$_x$.

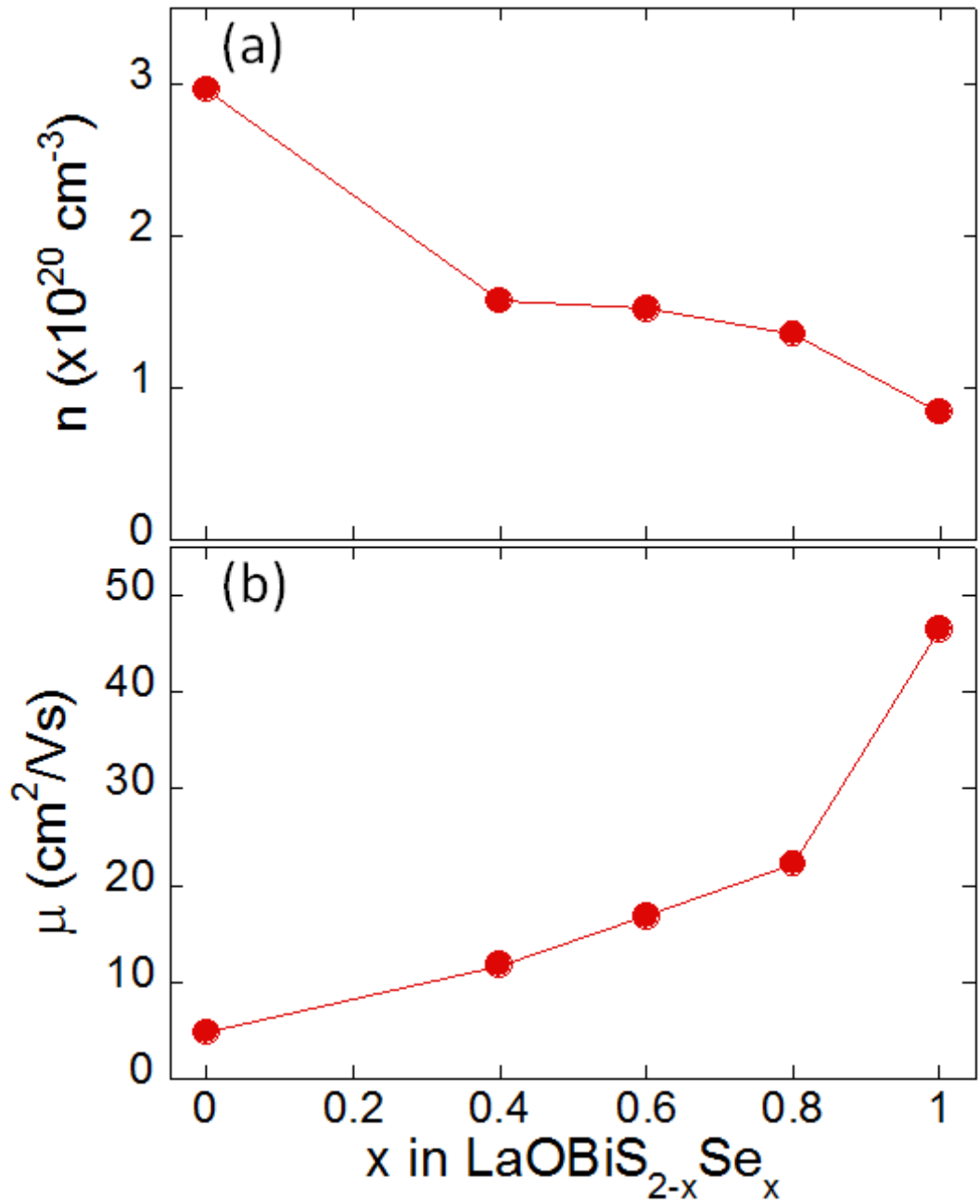

Fig. 2. Dependence of the (a) carrier concentration ($n$) and (b) carrier mobility ($\mu$) of densified LaOBiS$_{2-x}$Se$_x$ on the concentration ($x$) of selenium. The Hall measurements were performed at room temperature.